\documentclass[conference]{IEEEtran}
\usepackage{ifpdf}

\usepackage{cite}

\ifCLASSINFOpdf
   \usepackage[pdftex]{graphicx}
\else
   \usepackage[dvips]{graphicx}
\fi
\usepackage{amsmath}
\usepackage{algorithmic}

\usepackage{array}

\ifCLASSOPTIONcompsoc
  \usepackage[caption=false,font=normalsize,labelfont=sf,textfont=sf]{subfig}
  \else
  \usepackage[caption=false,font=footnotesize]{subfig}
\fi

\usepackage{dblfloatfix}
\usepackage{url}

\begin{document}
%
\title{Early prediction of the duration of protests using probabilistic Latent Dirichlet Allocation and Decision Trees}



%
\author{\IEEEauthorblockN{Satyakama Paul\IEEEauthorrefmark{1},
Madhur Hasija\IEEEauthorrefmark{2} and
Tshilidzi Marwala\IEEEauthorrefmark{3}}
\IEEEauthorblockA{\IEEEauthorrefmark{1}Oracle India Pvt. Ltd, Prestige Tech Park, Bengaluru, India - 560103 \\
Email: satyakama.paul@gmail.com}
\IEEEauthorblockA{\IEEEauthorrefmark{2}WNS Global Pvt. Ltd, Whitefield, Bengaluru, India - 560048\\
Email: madhur.hasija@gmail.com}
\IEEEauthorblockA{\IEEEauthorrefmark{3}University of Johannesburg, Johannesburg, South Africa - 2006\\
Email: tmarwala@uj.ac.za}}


\maketitle

\begin{abstract}
Protests and agitations are an integral part of every democratic civil society. In recent years, South Africa has seen a large increase in its protests. The objective of this paper is to provide an early prediction of the duration of protests from its free flowing English text description. Free flowing descriptions of the protests help us in capturing its various nuances such as multiple causes, courses of actions etc. Next we use a combination of unsupervised learning (topic modeling) and supervised learning (decision trees) to predict the duration of the protests. Our results show a high degree (close to 90$\%$) of accuracy in early prediction of the duration of protests. We expect the work to help police and other security services in planning and managing their resources in better handling protests in future.
\end{abstract}


%
\IEEEpeerreviewmaketitle

\section{Introduction}
Protests and agitations are an integral part of any democratic civil society. Not to be left behind when compared with the rest of the world, South Africa in recent years has also seen a massive increase in public protests. The causes of these protest were varied and have ranged from service delivery, labor related issues, crime, education, to environmental issues.\\

While in the past multiple studies and news articles have analyzed the nature and cause of such protests, this research uses a combination of unsupervised (topic modeling using probabilistic Latent Dirichlet Allocation (pLDA)) and supervised (single and ensemble decision trees) learning to predict the duration of future protests. We develop an approach in which an user inputting a description of a protest in free flowing English text, the system predicts the duration of the protest to a high (close to 90$\%$) degree of accuracy . We expect that an early correct prediction of the duration of the protest by the system will allow police and other security services to better plan and allocate resources to manage the protests.

\section{Problem Statement}

The objective of this research is to provide an early prediction of the duration of a protest based on South African protest data. The master dataset is obtained from the website - Code for South Africa \cite{IEEEhowto:CodeforSA}. It consisted of 20 features (columns) describing 876 instances (rows) of protests over the period of 1$^{st}$ February 2013 to 3$^{rd}$ March 2014. Among the 20 features, the statistically important and hence selected ones are shown in Table \ref{table:imp.stat.features} . The rest are repeated codification of the important features that convey the same statistical information as the important features. Hence they are ignored. Also detailed addresses of the location (Town or city name, First street, Cross street, Suburb area place name etc) of the protests are not considered. Instead the more accurate measures - Coordinates (latitudes and longitudes) are used.\\

\begin{table}[!t]
\renewcommand{\arraystretch}{1.3}
\caption{Important Statistical Features}
\label{table:imp.stat.features}
\centering
\begin{tabular}{|c|c|}
\hline
Whether Metro or not & Police station\\
\hline
Coordinates & Start date\\
\hline
End date & Cause of protest\\
\hline
Status of protest (Violent or not) & Reason for protest (text data)\\
\hline
\end{tabular}
\end{table}

\begin{table}[!t]
	\renewcommand{\arraystretch}{1.3}
	\caption{Percentage of protests vis-a-vis provinces}
	\label{table:prov.of.protests}
	\centering
	\begin{tabular}{|c|c|c|c|}
		\hline
		\textbf{Provinces} & \textbf{$\%$ of protests} & \textbf{Provinces} &\textbf{ $\%$ of protests}\\
		\hline
		\hline
		Gauteng & 37 & Limpopo & 6\\
		\hline
		Western Cape & 18 & Mpumalanga & 5\\
		\hline
		Kwazulu Natal & 14 & Free State & 3\\
		\hline
		Eastern Cape & 9 & Northern Cape & 2\\
		\hline
		North West & 6&&\\
		\hline
		
	\end{tabular}
\end{table}

Table \ref{table:prov.of.protests}, \ref{table:issues.of.protests}, \ref{table:state.of.protests} and \ref{table:duration.of.protests} shows the overall descriptive statistics of the protests during the above mentioned period. From the 876 rows, three are removed for which the one or more columns are missing\footnote{Since an insignificant percentage (0.34$\%$) of our data is missing we conveniently remove them without doing missing value imputation.}. Thus our modeling exercise is based upon 873 instances of protests. Table \ref{table:prov.of.protests} shows that Gauteng\footnote{Gauteng has in it Johannesburg and Pretoria. Johannesburg is the commercial hub of South Africa, and the office of President and Cabinet is in Pretoria.} as the seat of commerce, and President and Cabinet; and Western Cape\footnote{Cape Town as the seat of South African parliament is situated in Western Cape.} as the legislative capital has the largest concentration of protests, followed by the others. Table \ref{table:issues.of.protests} shows that the three largest issues of protests are from service delivery, labor related issues, and crime related at respectively 31$\%$, 30$\%$ and 12$\%$ of the total. From table \ref{table:state.of.protests} it can be seen that the difference between peaceful and violent protests is low at 55$\%$ and 45$\%$ respectively. However the most interesting insight comes from the duration of protests\footnote{Duration of protests is the difference between the End day and the Start Day of the protest.}. From table \ref{table:duration.of.protests} it can be observed that the majority of protests (at 74.34$\%$ of the total) last for less than 24 hours. Thus this feature is highly skewed. \\

\begin{table}[!t]
	\renewcommand{\arraystretch}{1}
	\caption{Percentage of protests vis-a-vis issues}
	\label{table:issues.of.protests}
	\centering
	\begin{tabular}{|c|c|c|c|}
		\hline
		\textbf{Issue} & \textbf{$\%$ of protests} & \textbf{Issue} &\textbf{ $\%$ of protests}\\
		\hline
		\hline
		Service delivery & 31 & Political  & 4\\
		\hline
		Labour  & 30 & Transport  & 3\\
		\hline
		Crime  & 12 & Xenophobia & 2\\
		\hline
		Election  & 6 & Individual causes & 1\\
		\hline
		Vigilantism & 5 & Environment  &1\\
		\hline
		Education   &5 &&\\
		
		\hline
		
	\end{tabular}
\end{table}

\begin{table}[!t]
	\renewcommand{\arraystretch}{1.3}
	\caption{Percentage of protests vis-a-vis state }
	\label{table:state.of.protests}
	\centering
	\begin{tabular}{|c|c|}
		\hline
		
		\textbf{State} & \textbf{$\%$ of protests} \\
		\hline
		\hline
		Peaceful & 55\\
		\hline
		Violent & 45\\
		\hline
		
	\end{tabular}
\end{table}

\begin{table}[!t]
	\renewcommand{\arraystretch}{1.3}
	\caption{Percentage of protests vis-a-vis duration in days }
	\label{table:duration.of.protests}
	\centering
	\begin{tabular}{|c|c|}
		\hline
		\textbf{Duration} & \textbf{$\%$ of protests} \\
		\hline
		\hline
		0 (less than 24 hrs) & 74.34\\
		\hline
		1 & 11.34\\
		\hline
		2 & 4.58\\
		\hline
		3 & 2.06\\
		\hline
		4 & 2.17\\
		\hline
		5-13,19, 21-23, 31,  34, 37, 39,   57, 65 & less than 1$\%$\\
		\hline
		
	\end{tabular}
\end{table}

\begin{figure}[!t]
\centering
\includegraphics[width=3.5in]{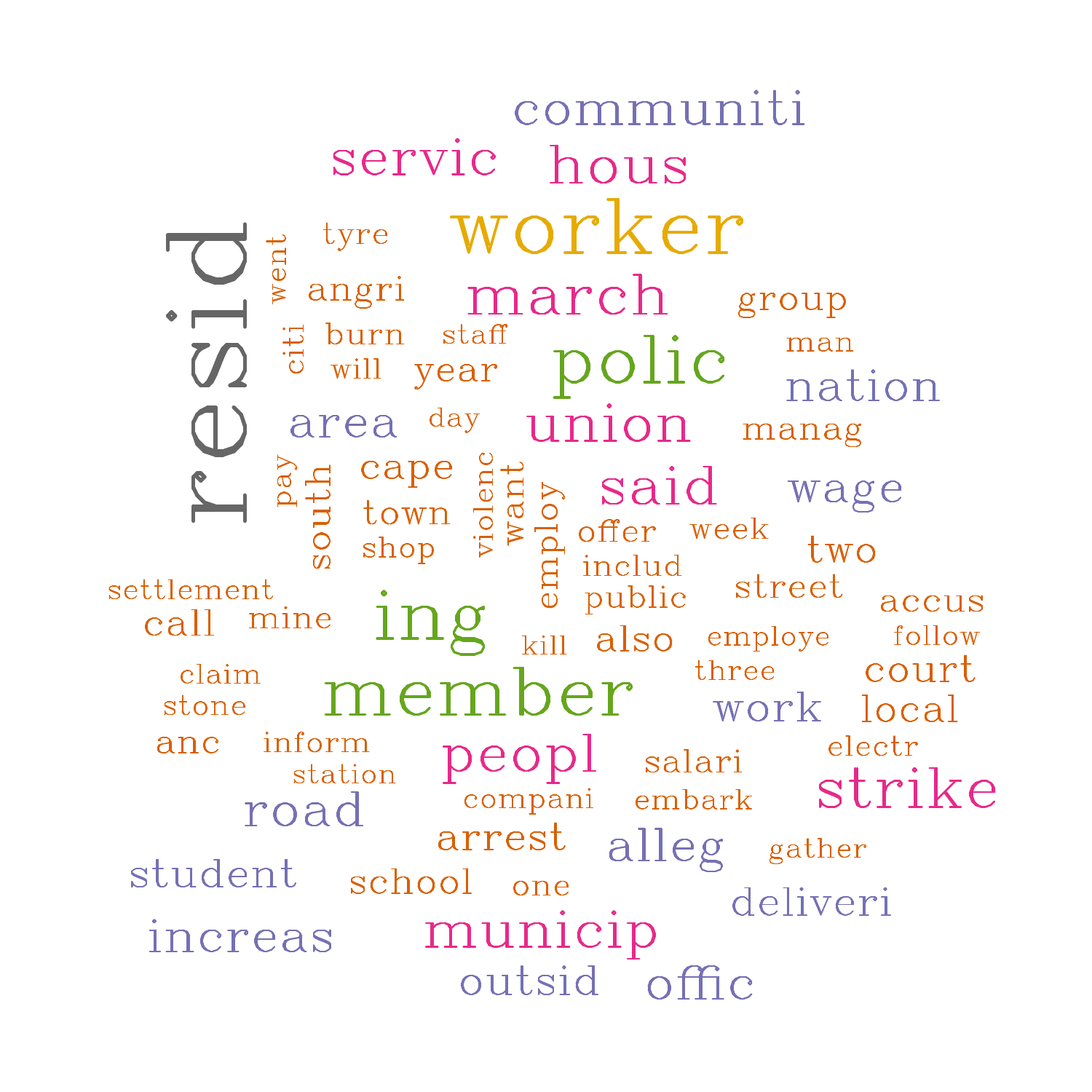}
\DeclareGraphicsExtensions.
\caption{Word cloud of the entire text courpus of the protests}
\label{figure:wordcloud}
\end{figure}

The idea behind this work is to use only the text description of the protests (predictor variable) to predict the duration of future protest(s) (response variable). Some typical examples of the descriptions of protests are as follows. Flagged
as service delivery and violent - ``Residents of both towns Butterworths and Centane blockaded the R-47 between the two towns, accusing the Mnquma Municipality of ignoring
their request for repairs to the road.'' Eyeballing the text does not indicate any violence. A second example flagged as service delivery and peaceful protest is ``ANGRY community leaders in four North West villages under the Royal Bafokeng Nations jurisdiction protested this week against poor services and widespread unemployment among the youth. Now they not only demand their land back, but want a 30$\%$ stake in
the mines which are said to employ labour from outside the villages.''. While the first line of the text referred to the cause of the protest as service delivery, the second line referred to political (demand for return of land) and labor issues (3$\%$
stake in mines and corresponding increase in employment). In this sense, we believe that strict flagging of protests into one category or another restricts the knowledge of the protests. We also believe that it is normal for human social concerns to
spill from one area to another during protests that is not well captured by only one restrictive flag attached to one protest. Thus we drop the categorical features (with strict class labels) and consider only the text descriptions of the protests as a
predictor of its duration. Another important advantage of using text descriptions is that they give flow/progress of events that occurred during a protest and other relevant details. Figure \ref{figure:wordcloud} shows the word cloud of the entire protest corpus. Prior building the word cloud the usual preprocessing on the text corpus such as removal of punctuation, numbers, common English stopwords, white spaces etc have been carried out.
The cloud consists of 75 words\footnote{The number of words is kept low for better visibility.} and words that occur with a minimum of at least 25 times are included. The more the frequency of the words occurring in the text corpus, the bigger is its font size. Lastly due to word stemming, “resid” is created from words like residence and residing.

\section{THEORETICAL CONCEPT - PROBABILISTIC LATENT DIRICHLET ALLOCATION}

In this subsection we provide a short introduction to pLDA. LDA is an unsupervised generative probabilistic model primarily used for topic discovery. It states that for a collection of words in a large number of documents, each document is a mixture of a number of ``latent topics'' \footnote{Note that the topics are latent because they are neither defined semantically nor epistemologically.} and each word in a document is a result of a latent topic. Following \cite{IEEEhowto:BleiNgJordan2003}, \cite{IEEEhowto:Reed}; a document is a random mixture of latent topics and each topic\footnote{Topic and latent topic are the same and used interchangeably} in turn is created by the distribution of words. Mathematically the LDA model can be stated as follows. For each document \textbf{w} in a corpus \textit{D}

\begin{enumerate}
	\item Choose $N \sim Poison(\xi)$
	\item Choose $\theta \sim Dir(\alpha)$
	\item For each of the \textit{N} words $w_{n}$:
		\begin{enumerate}
			\item Choose a topic $w_{n} \sim Multinomial(\theta)$
			\item Choose a word $w_{n}$ from $p(w_{n}|z_{n}, \beta)$, a multinomial probability conditioned on the topic $z_{n}$
		\end{enumerate}
\end{enumerate}

where a document \textbf{w} is a cobination of \textit{N} words, for example $\textbf{w} = (w_{1}, w_{2}, ...., w_{N})$. A corpus \textit{D} is a collection of \textit{M} documents such that $D = (\textbf{w}_{1}, \textbf{w}_{2}, ...., \textbf{w}_{M})$. $\alpha$ and $\beta$ are
parameters of the Dirichlet prior on per document topic and
per topic word distribution respectively. \textit{z} is a vector of topics.
The central idea of LDA is to find the posterior distribution
($\theta$) of the topics (z) when the document (\textbf{w}) is given, i.e.

\begin{equation}
p(\theta, z|\textbf{w}, \alpha, \beta) = \dfrac{p(\theta, z, \textbf{w}|\alpha, \beta)}{p(\textbf{w}|\alpha, \beta)}
\end{equation}

Since it is beyond the scope of this paper to derive the
detailed formula, we summarize the two other important
results that will subsequently be required in our analysis. The
marginal distribution of a document is:

\begin{equation}
p(\textbf{w}|\alpha, \beta) = \int p(\theta|\alpha)(\prod_{n=1}^{N})\sum p(z_{n}|\theta)p(w_{n}|z_{n}, \beta)d\theta
\end{equation}

The probability of a corpus is:

\begin{equation}
p(D|\alpha, \beta) = \prod_{d=1}^{m}\sum p(\theta_{d}|\alpha)(\prod_{n=1}^{N_{d}}p(w_{dn}|z_{dn}, \beta)d\theta)
\end{equation}

\section{Experimental Setup}

\begin{figure}[!t]
	\centering
	\includegraphics[width=3.5in]{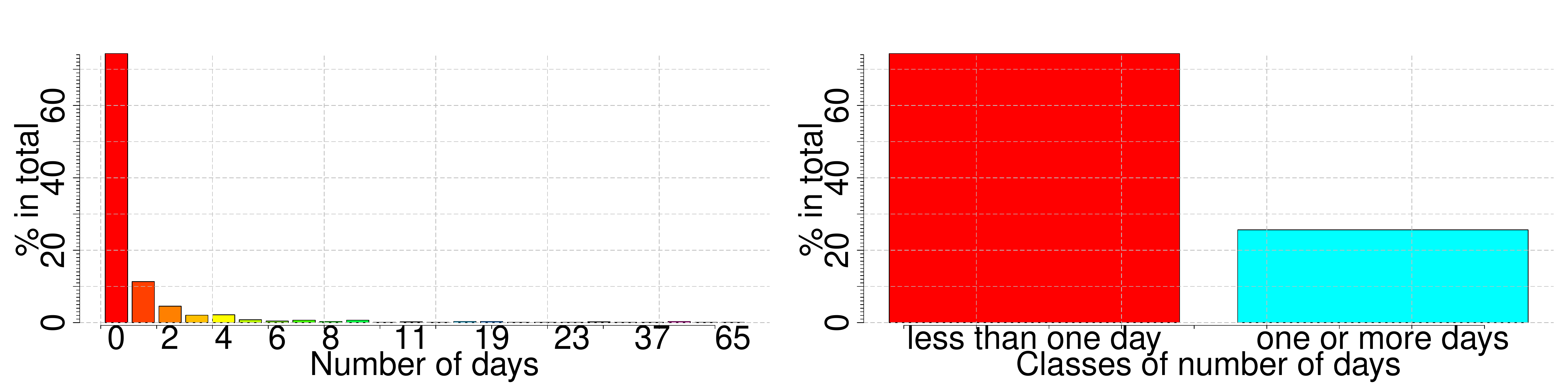}
	\DeclareGraphicsExtensions.
	\caption{Setting of two class classification problem}
	\label{figure:twoclass}
\end{figure}

As seen in table \ref{table:duration.of.protests}, our response variable - duration of protests is highly schewed. Protests lasting for less than 24 hours is 74.34$\%$ of the total number of protests. Protests above one day is less than 5$\%$ of the total
number. In effect it means that in reality, South Africa rarely experiences protests that stretches beyond one day. So for practical purposes, we couple protests lasting above 24 hours into one class and compare it against protests lasting below
24 hours. The right panel of figure \ref{figure:twoclass} shows the percentage of protests falling in less than one day (74.34$\%$) and one or more days (25.66$\%$). Thus the binary classification problem now is to correctly predict the response variable (less than one day against one or more days) for the text corpus of the protest descriptions.\\

However before getting into the classification exercise we need to perform two tasks. One, we need to find the optimal number of latent topics from the text corpus. Two, since classification algorithms per se cannot take text documents, we need to extract a set of latent topics for each text description of a protest. In the next section we discuss the above tasks and their results.

\section{RESULTS AND DISCUSSIONS}

\begin{figure}[!t]
	\centering
	\includegraphics[width=3.5in]{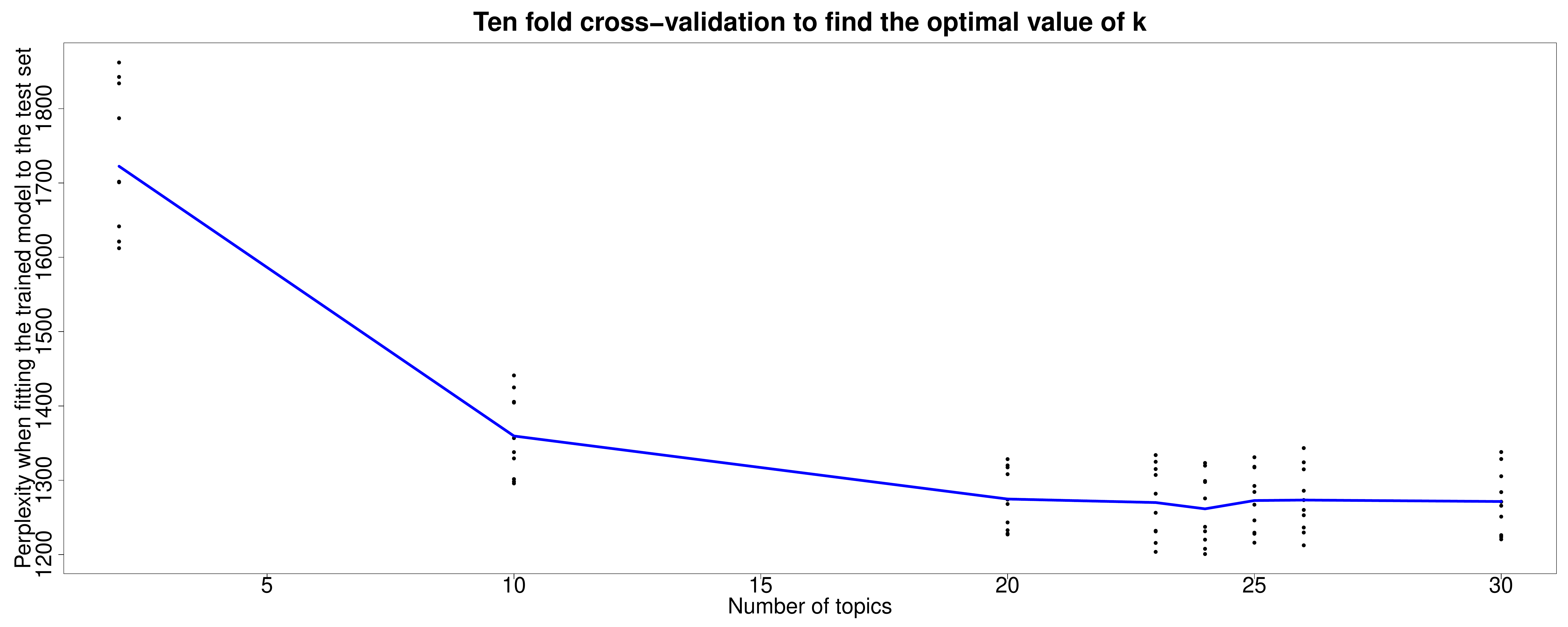}
	\DeclareGraphicsExtensions.
	\caption{Finding the optimal number of topics}
	\label{figure:opt.no.topics}
\end{figure}

Finding the optimal number of hidden topics from the text corpus is an important task. If the chosen number of topics is too low, then the LDA model is unable to identify the accurate classifiers. However if the number is too high, the model becomes increasingly complex and thus less interpretable \cite{IEEEhowto:Zhao2015} \cite{IEEEhowto:Zhao2014}. In contrast to the often used procedure of intelligently guessing the optimal number of topics in a text corpus, following \cite{IEEEhowto:BleiNgJordan2003} we use the perplexity approach to find the optimal number of topics. Often used in language/text based models, perplexity is a measure of ``on average how many different equally most probable words can follow any given word'' \cite{IEEEhowto:NClab}. In other words its a measure of how well a statistical model (in our case LDA) describes a dataset \cite{IEEEhowto:Zhao2015} where a lower perplexity denotes a better model. It is mathematically denoted by:

\begin{equation}
perplexity(D_{test}) = exp \{- \dfrac{\sum_{d=1}^{M}logp(\textbf{w}_{d})}{\sum_{d=1}^{M} N_{d}} \}
\end{equation}

where the symbols have the same meaning as in section III. Using a trial number of topics that range between 2 to 30\footnote{x axis of fig \ref{figure:opt.no.topics}  shows the trial number of topics}, we use perplexity to find the optimal number of topics from our text corpus. The text corpus is broken into a training and test set. A ten fold cross validation is carried out using 1000 iterations. The black dots show the perplexity score for each fold of cross validation on the test set for various numbers of topics. The average perplexity score is shown by the blue line in figure \ref{figure:opt.no.topics} and the score is least for 24 latent topics. Thus the optimal number of topics for our text corpus is 24.\\

Next we perform a LDA with 24 topics on our text corpus. Table \ref{table:text.desc.assoc.topic.prob}  shows an example of a text description and the probabilities associated with the various topics.\\

\begin{table}[!t]
	\renewcommand{\arraystretch}{1.3}
	\caption{AN EXAMPLE OF A TEXT DESCRIPTION AND ITS ASSOCIATED TOPIC PROBABILITIES}
	\label{table:text.desc.assoc.topic.prob}
	\centering
	\begin{tabular}{|c|c|c|c|c|c|}
		\hline
		
		\textbf{Text} & \textbf{P(T$_{1}$)..} & \textbf{..P(T$_{8}$)} &\textbf{...} &\textbf{..P(T$_{20}$)} &\textbf{..P(T$_{24}$)} \\
		\hline
		
		The residents wanted &0.0027&0.15&...&0.79&0.0028\\
		Sterkspruit to be  &&&&&\\
				moved from the &&&&&\\
		Senqu municipality &&&&&\\
		and be a municipality &&&&&\\
		on its own  &&&&&\\
		
		\hline
		
	\end{tabular}
\begin{flushleft}
	*** \textbf{P(T$_{n}$)} refers to probability of the n$^{th}$ topic where n varies from 1 to 24. Here T$_{1}$ = shop, T$_{8}$ = march, T$_{20}$ = municip and T$_{24}$ = anc.
	
\end{flushleft}
\end{table}

The topic names are given in the bottom of the table. The topics court and resid have the highest probabilities for the text. It might also be noted that since the topics are multinomial distributed to the entire text corpus and the words in corpus, hence there is no direct or visible relationship
that connects the topics with the text descriptions. We can only assume that the topics are complexly related to the text descriptions.\\

Table \ref{table:text.desc.assoc.highest.topic.prob} shows the topics with four largest probabilities (largest, 2nd largest,... 4th largest) associated with a text description. With computational concerns in mind, we restrict the largest probabilities to only four.\\

\begin{table}[!t]
	\renewcommand{\arraystretch}{1.3}
	\caption{AN EXAMPLE OF TEXT DESCRIPTION AND ITS ASSOCIATED HIGHEST PROBABILITY TOPICS}
	\label{table:text.desc.assoc.highest.topic.prob}
	\centering
	\begin{tabular}{|c|c|c|c|c|}
		\hline
		
		\textbf{Text} & \textbf{LP} & \textbf{2$^{nd}$ LP}  &\textbf{3$^{rd}$ LP} &\textbf{4$^{th}$ LP} \\
		\hline
		
		The residents wanted &municip&march&resid&hospit\\
		Sterkspruit to be  &&&&\\
		moved from the &&&&\\
		Senqu municipality &&&&\\
		and be a municipality &&&&\\
		on its own  &&&&\\
		
		\hline
		
	\end{tabular}
	\begin{flushleft}
		*** \textbf{P(T$_{n}$)} refers to probability of the n$^{th}$ topic where n varies from 1 to 24. Here T$_{1}$ = shop, T$_{8}$ = march, T$_{20}$ = municip and T$_{24}$ = anc.
		
	\end{flushleft}
\end{table}

It might be recollected from the last paragraph of section IV where we stated that classification algorithms per se cannot take text documents as a predictor variable. Table \ref{table:text.desc.assoc.highest.topic.prob} shows a way in which a single text description can be represented as a set of most relevant\footnote{The most relevant topics are the ones with highest probabilities.} topics. Thus for an entire corpus of 873 text documents, the predictor side of the classification model would consist of a topic matrix of 873x4 dimensions. Including the response variable, our modeling data would have 873x5 dimensions.\\

In addition, it might also be recalled that our response variable duration of protest is unbalanced with the percentage of values falling in the class - less than one day at 74.34$\%$ and one or more days at 25.66$\%$. Since unbalanced classes are not learned well by decision trees, so we use both way balanced sampling strategies to balance the modeling data. Next the modeling data is split into a training and test set in the 7:3 ratio. The dimensions of the training and test sets are 910x5 and 388x5 respectively.\\

In the next step we use decision tree based classification algorithms to model the relationship between the topic matrix and the duration of the protests. Specifically we use three algorithms - C 5.0, treebag, and random forest and do a ten fold cross validation with five repeated iterations. The prediction results for the class one or more days on the test dataset is shown in table \ref{table:perf.metric}.\\

\begin{table}[!t]
	\renewcommand{\arraystretch}{1.3}
	\caption{PERFORMANCE METRICS OF VARIOUS ALGORITHMS ON THE TEST SET}
	\label{table:perf.metric}
	\centering
	\begin{tabular}{|c|c|c|c|}
		\hline
		
		\textbf{} & \textbf{C5.0} & \textbf{Treebag}  &\textbf{Random forest}  \\
		\hline

		Balanced accuracy &79.38$\%$ &88.40$\%$ &89.69$\%$\\
		\hline
		Kappa  &0.590 &0.769 & 0.795\\
		\hline
		Sensitivity &87.03$\%$ &94.59$\%$ &95.68$\%$\\
		\hline
		Specificity &72.41$\%$ &82.76$\%$ &84.24$\%$ \\

		\hline
		
	\end{tabular}
	\begin{flushleft}
		*** \textbf{P(T$_{n}$)} refers to probability of the n$^{th}$ topic where n varies from 1 to 24. Here T$_{1}$ = shop, T$_{8}$ = march, T$_{20}$ = municip and T$_{24}$ = anc.
		
	\end{flushleft}
\end{table}

Treebag with a number of decision trees is able to better predict the response variable than a single tree in C5.0 because multiple trees help in reducing variance without increasing bias. Again, random forest performs better than treebag because in addition to multiple trees it also randomly
selects a subset of the total number of features at each node. Further best split feature from the subset is used to split each node of the tree. This salient feature of random forest is absent
in treebag. The combination of a number of trees and random selection of a subset of features at each node of the tree helps in further reducing the variance of the model. Thus in effect, random forest performs best in predicting the duration of the protests on the test dataset.\\

The codes for the analysis are in \texttt{\url{https://www.dropbox.com/s/dzyj2lviqnlgk5x/dropbox_ieee_la_cci_2017.zip?dl=0}}

\section{LIMITATIONS}

The first limitation of this research is that LDA does not allow for evolution of topics over time. This means that if the nature and scope of the protests remain pretty stagnant over time then our model is expected to perform fairly well. Second, since it is a bag of words model, so sentence structures are not considered. And third, topics are not independent of one another. Thus it might happen that the same word represents two topics creating a problem of interpretability.

\section{Conclusion}

This paper is an combination of unsupervised and supervised learning to predict the duration of protests in South African context. Protests and agitations being social issues; have multiple nuances in terms of causes, courses of actions etc that cannot be very well captured by restrictive tags. Thus we discard the approach of restrictive characterization of protests and use free flowing English texts to understand its nature, cause(s), course of action(s) etc. Topic discovery (unsupervised learning using pLDA) and subsequent classification (supervised learning using various decision tree algorithms) provides promising results in predicting the duration of protests. We expect that the implementation of the framework can help police and other security services in better allocating resources to manage the protests in future.


\begin{thebibliography}{1}

  
\bibitem{IEEEhowto:CodeforSA}
Code for South Africa, \emph{Protest Data},\hskip 0.5em plus
0em minus 0em\relax \url{https://data.code4sa.org/dataset/
Protest-Data/7y3u-atvk}.  

\bibitem{IEEEhowto:BleiNgJordan2003}
D.M. Blei, A.Y Ng and M.I Jordan, \emph{Latent Dirichlet Allocation},\hskip 0.5em plus
0em minus 0em\relax Journal
of Machine Learning Research, Vol. 3, pp.993-1022, 2003.


\bibitem{IEEEhowto:Reed}
C. Reed, \emph{Latent Dirichlet Allocation: Towards a Deeper Understanding},\hskip 0.5em plus
0em minus 0em\relax \url{http://obphio.us/pdfs/lda tutorial.pdf}.


\bibitem{IEEEhowto:Zhao2015}
W. Zhao, J.J Chen, R. Perkins, Z. Liu, W. Ge, Y. Ding and W. Zou, \emph {A heuristic approach to determine an appropriate number of topics in topic modeling},\hskip 0.5em plus 0em minus 0em\relax 12th Annual MCBIOS Conference, March, 2015.


\bibitem{IEEEhowto:Zhao2014}
W. Zhao, W. Zou, J.J Chen, \emph {Topic modeling for cluster analysis of large biological and medical datasets},\hskip 0.5em plus 0em minus 0em\relax BMC Bioinformatics, Vol. 15 (Suppl 11), 2014.


\bibitem{IEEEhowto:NClab}
NCLab, \emph {Perplexity},\hskip 0.5em plus 0em minus 0em\relax \url{http://nclab.kaist.ac.kr/twpark/htkbook/node218_ct.html}.


%
%



\end{thebibliography}
\end{document}